\documentclass[12pt]{article}
\usepackage{color}
\usepackage{graphicx}

\definecolor{violet}{rgb}{0.4,0,0.6}
\definecolor{vert}{rgb}{0,0.4,0.2}
\definecolor{navy}{rgb}{0.0,0.0,0.4}

\def\spose#1{\hbox to 0pt{#1\hss}}\def\lta{\mathrel{\spose{\lower 3pt\hbox
{$\mathchar"218$}}\raise 2.0pt\hbox{$\mathchar"13C$}}}  \def\gta{\mathrel
{\spose{\lower 3pt\hbox{$\mathchar"218$}}\raise 2.0pt\hbox{$\mathchar"13E$}}}

\def\be{\begin{equation} } \def\fe{\end{equation}}

\begin{document}
\title{Anthropic interpretation of quantum theory}

 \author{Brandon Carter\\
 LuTh, Observatoire de Paris, 92195 Meudon } 
  
\date{July, 2003}

\maketitle

\vskip 1.2 cm

\noindent
{\bf Abstract. }
 The problem of interpreting quantum theory on a large (e.g. cosmological) 
scale has been commonly conceived as a search for objective reality in a 
framework that is fundamentally probabilistic. The Everett programme 
attempts to evade the issue by the reintroduction of determinism at the 
global level of a ``state vector of the universe''. The present approach is 
based on the recognition that, like determinism,  objective reality is an 
unrealistic objective. It is shown how  an objective  theory of an 
essentially subjective reality can be set up using an appropriately
weighted probability measure on the relevant set of Hilbert subspaces. It 
is suggested that an entropy principle (superseding the weak anthropic 
principle) should be used to provide the weighting that is needed.

\vskip 2 cm

{\it Text for exposition presented at Interdisciplinary Colloquium
``La Philosophie de la Nature aujourd'hui?'', Paris, March 2003,
and 8th Peyresq Physics Meeting ``The Early Universe'', June 2003.}


\vfill\eject

\noindent\textit{ Introduction}

Among the diverse schools of thought concerning the interpretation of 
quantum theory, it is the one founded by Everett~\cite{Everett} that can 
claim the widest nominal adherence~\cite{TegmarkWheeler} (if not an 
absolute majority, surely more than for any specific alternative) within 
the theoretical physics community. Those who are not satisfied by the 
Everett doctrine belong to two distinct groups. The first (including 
perhaps a majority of experimental, as opposed to theoretical, physicists) 
consists of those who have not thought about the issue, and who are 
content with earlier Copenhagen type interpretations such as are useful in 
a restricted laboratory context but inapplicable in a broader, e.g. 
cosmological, context. The others, namely those who are not so content 
(and who have often been inclined to search -- vainly -- for acceptable 
alternatives of a deterministic nature) are of two kinds, of which perhaps 
the most numerous consists of those who dislike the prolificity implicit 
in Everett's ``many world'' idea. The remaining minority consists of those 
who have looked into the matter sufficiently carefully to be aware of the 
intrinsic intellectual deficiency of the Everett doctrine.

In order to satisfy the requirements of the latter group, namely those who 
are broad minded about ontology, but more rigourously demanding in so far 
as intellectual coherence is concerned, the alternative interpretation 
proposed here is based on a rather different approach that incorporates 
ideas developed particularly by Dyson~\cite{Dyson} and by Page~\cite{Page}.

\medskip
\noindent\textit{Deficiency of the Everett interpretation}

It has been recognised on many occasions that Everett's doctrine is 
intrinsically self contradictory if taken too literally, while on the other 
hand, if taken less literally, the Everett  ``interpretation'' is so ambiguous 
as to be essentially meaningless unless it is itself provided with an 
appropriately coherent interpretation, of which various 
kinds~\cite{Barrett} have been envisaged 

As in more traditional approaches, the starting point of the Everett 
doctrine is the generally accepted principle that observational 
discrimination is describable in quantum theory as a choice between {\it 
mutually orthogonal Hilbert subspaces}, which are describable as the 
alternative {\it branch-channels}, each of which is characterised by a 
corresponding Hilbert space projection operator ${\bf e}={\bf e}^2$.

It is to be noted that many writings, including those of Everett, convey 
the potentially misleading idea that the number of such ``branches'' will 
automatically multiply as successive observations are made. To see that,
as emphasised by Tippler~\cite{Tippler}, this need not -- and indeed, beyond 
a certain point, ultimately cannot -- be the case, it suffices to
consider the usual kind of toy sensor model, in which what is observed is a 
combination of up or down spin orientations (with respect, for example, to 
a background magnetic field direction) for a set of let us say $S$ sensor
particles.   For a single particle initially in a pure -- let us say down --
state, an elementary measurement will indeed double the number of relevant 
branch channels, but such doubling can give rise to at most a maximum value 
${\cal N}=2^S$ of branch-channels, corresponding to the recording of a 
maximum of $S$ bits of information. Subsequent observations can record new 
information only at the expense of {\it erasure} (by conversion~\cite{Vedral}
into Landauer entropy) of part of what was previously recorded  -- unless of 
course the number $S$ is itself increased. Extrapolating from the example 
of such a toy to the case of a human observer, it is to be remarked that 
the relevant information capacity $S$ will indeed increase progressively 
during infancy (though not quite fast enough to prevent children from 
forgetting as well as learning) but that it will saturate during 
adulthood (and ultimately decrease if a stage of senility is attained).

Whether their number is increasing or not, a set of probabilities for such 
alternative branch-channels is supposed, in the traditional approach, to be 
provided by a prescription whereby any particular branch channel, 
$e$ say, will have probability
\be P_{[{\cal O}]}\{e\}={\rm tr}\{ {\bf P}\,{\bf e}\}\, ,\label{1}\fe
given by the appropriate choice of a von Neumann type (unit trace) operator, 
${\bf P}$, which (as in the classical Bayesian case) is readjusted to take 
account of new observational information as and when it becomes available. 
It is to be noted that this is just a conditional probability, subject to 
the requirement that the relevant observation, ${\cal O}_e$ say, be actually 
carried out. Further information must be supplied (as discussed below) if 
one needs an absolute probability $P\{e\}$  allowing for the unlikelihood 
of that particular observation ever really being performed.

This works very well in a laboratory context for an observer {\it outside} 
the system under consideration, but in more general (e.g. cosmological) 
contexts, involving more than one observer {\it inside} the system, the 
desideratum of using a probability distribution that treats distinct 
observers (having access to different information) on the same objective 
footing poses a problem. 

The Everett strategy for dealing with the problem of the subjectivity
of observational readjustment of the probability distribution is to deny 
any objective significance for observational discrimination by 
proclaiming~\cite{Everett} that {\it all} the relevant branch-channels 
are ``equally real''. The trouble with this is  with this is that, if 
taken at its face value, it disagrees with ordinary laboratory experience, 
for which the probability interpretation works so well. In what Graham 
describes~\cite{Graham} as an attempt to to ``escape from this dilemma'' 
Everett is obliged to resort to an Orwellian gambit whereby the meaning of 
of ``reality'', and of ``equality'', is cast into doubt by the admission 
that the branch-channels thus described are nevertheless characterised by 
{\it different} ``weightings''. Although they are supposed to be given 
in the usual way by a unit trace operator ${\bf P}$, these ``weightings'' 
are deemed not to be ordinary subjective probabilities but to have an 
objective physical status inherited from a physically ``real'' state vector 
$\vert\Psi\rangle$ of the universe, in terms of which the ``weighting'' 
operator has the pure state form ${\bf P}=\vert\Psi\rangle\langle\Psi\vert$. 

Opinions may differ about the most appropriate choice of the cosmological 
state function~\cite{HH,Vilenkin} and about whether the ontological 
prolificity entailed  by the ``many worlds'' idea is appealing or 
appaling~\cite{Leslie}, but what is intrinsicly wrong with Everett's so 
called ``interpretation'' is his failure to provide a self-consistent 
interpretation of the terms -- such as ``weighting'' and particularly 
``reality'' -- on which his formulation relies.

The approach described below provides a remedy in which the concept of 
``reality'' is interpreted in a more realistically restrictive manner,
while the ``weightings'' are reinstated in their traditional role as 
(conditional) probabilities in the usual (Born) sense.

\medskip
\noindent\textit{A viable approach}

Although other kinds of reality can be imagined (e.g. by theologians) 
the only reality of which we have a direct knowledge is that
of a {\it subjective mental perception}, which can of course include 
memories of previous perceptions, which -- as discussed by Page~\cite{Page}
-- are collectively construed as mind states of the perceptor. Many such 
mind states are interpretable merely as dreams, but many more are 
interpretable as corresponding to events in an extrinsic physical world. 
The aim of the natural (physical and biological) sciences is to understand 
how this physical world works and {\it  how our perceptions fit into it}. 
These sciences provide descriptions of various aspects of the extrinsic 
world in terms of more or less elaborate theoretical models in which events 
that are actually supposed to occur are selected within a wider class of 
{\it eventualities} that may or may not occur. (For example a kind of toy 
model that is particularly popular for pedagogical and other purposes in 
theoretical physics consists of a scalar field governed by a hyperbolic 
differential equation whose solutions are the eventualities, of which a 
particular member is selected as an actual event by some specific choice of 
initial conditions.)

It is sometimes useful to employ a model of the simplified kind
(exemplified by that of the traditional criminal jury, for which the 
only admissible eventualities are ``guilty'' and ``not guilty'') that 
is qualifiable as deterministic, meaning that the specification  of the 
events that actually occur is provided in an unambiguous manner. However 
it is often more realistic to employ a model that is {\it indeterminate}, 
in the sense that it leaves the question of which eventualities occur as 
actual events to be decided (or not, as the case may be) by a process of 
observation and perception that may (subsequently) take place {\it 
outside} the framework of the model under consideration. In the most 
effective kind of indeterminate model, a guide to what is most likely to 
be (subsequently) observed is provided by the attribution to the relevant 
eventualities of a {\it probability} weighting that is conceived (there 
is a vast literature on this subject) as a generalisation of the relative 
frequency that would be obtained if (as is seldom the case) it were 
possible perform an unlimited number of observation and perception 
processes under identical conditions. 

Among such probabilistic models, a particularly important category is that 
of {\it quantum} theoretical models, which provide the most fundamental 
kind of physical description available today. This means that other 
deterministic or probabilistic descriptions (such as were sufficient for 
describing what was known before the end of the nineteenth century) can be 
considered just as approximations to an underlying quantum description 
(which at some future epoch may itself turn out to be qualifiable as an 
approximation to something even more mysterious that has not yet been 
conceived, and for which there is as yet no obvious necessity). In a 
quantum theory, an admissible eventuality, $e$ say, is supposed to be 
identifiable with a corresponding Hilbert subspace, for which a  conditional 
probability weighting is provided by a von Neumann type operator, ${\bf P}$ 
say, according to a prescription of the form (\ref{1}).

Since the various conceivable eventalities will in general not be mutually 
exclusive, the sum of their conditional probabilities $P_{[{\cal O}]}\{e\}$ 
can greatly exceed unity. A well behaved probability measure $P\{e\}$ on the 
set of Hilbert subspaces will however  be obtainable if a probability 
$P\{ {\cal O}_e\}$ for the actual occurrence of a corresponding {\it 
observational discrimination} (between $e$ and a set of {\it mutually 
orthognal} alternatives that together form an ``observable'' set completely 
spanning the Hilbert space of the system) is provided within a suitably 
extended framework that includes an appropriate {\it sensor} system. The 
probability $P\{e\}$ for the eventuality $e$ to be actually observed will 
then be given by 
\be P\{e\}=P\{{\cal O}_e\}\, P_{[{\cal O}]}\{e\}\, .\label{2}\fe
Of course for most of the so called ``observables'' in physical models, such 
an actual observation would be practically impossible or at best highly 
improbable, so that the relevant factor $P\{{\cal O}_e\}$  would be 
negligibly small.
 
In order to use an abstract physical model as a guide to ``reality'',
meaning~\cite{Page} what is actually perceived, it is of course necessary
to postulate that there should be a correspondence between the {\it mind
of the perceptor} and the {\it brain of a physical observer}. It is to be 
understood that the latter is identifiable within the model as a physical 
sensor subsystem, for which there is a special subclass of eventualities -- 
called {\it perceptibles} -- of the particular kind to which the the only 
events that are undisputably real -- namely those identifiable with 
{\it actual perceptions} -- must belong.

In order to avoid attribution of a privileged status to some particular 
physical observer such as Schroedinger (the performer of the famous cat 
experiment) or (like many a subconscious solipsist) oneself, it is 
necessary to situate the probabilistic (quantum or even classical) physical 
model within the framework of an extended perceptual system in which the 
perceptor has the role of an ultimate sensor of sensors. The eventualities 
within this perceptual system consist, not just of those within the 
underlying quantum mechanical system, but in addition, for each perceptible 
eventuality $e$ there  is a corresponding perceptual eventuality, 
${\cal O}_e$ say, namely that of being being subject to an actual 
observational discrimination by the perceptor. To complete the structure 
of the perceptual theory,  all that remains is to specify a corresponding 
distribution of probabilities  $P\{{\cal O}_e\}$ (that must vanish for 
non-perceptible eventualities) in a manner that avoids giving {\it a
priori} privilege to one's own situation.
 
When this is done, the probability, $P\{e\}$ say, for a particular perception 
to be the one corresponding to the physical eventuality $e$ will be given by 
an expression of the form (\ref{2}) in which  $P_{[{\cal O}]}\{e\}$ is the 
ordinary physical probability as given, according to (\ref{1}) by the 
underlying quantum mechanical model, while the factor $P\{{\cal O}_e\}$ is 
something that must be provided by an ansatz that appropriately refines the 
original (weak) anthropic principle~\cite{Carter,Bostrom} whereby equal 
weight was attributed to comparable anthropic observers.

\medskip
\noindent\textit{Entropic principle.}

For the present purpose it is necessary not just to attribute a
probability weighting to the entire life of an observer qualifiable 
as anthropic (in the sense of being sufficiently similar to ourselves) 
but to individual perceptions in the lives of observers who may be very 
different. It is to be reiterated that these perceptions are not just 
simple events (like the recognition that one's cat is or is not still 
alive) but that they may involve an intricate web of memories 
corresponding to something at least as complicated as what are known 
as {\it consistent histories}~\cite{GH}.
 
A first idea that comes to mind is to take the required weighting  
$P\{{\cal O}_e\}$ for such a perception to be proportional to its proper 
time duration, which in our human case is generally considered~\cite{Page} 
to be of the order of a fraction of a second. However the line of 
reasonning developped in Dyson's discussion~\cite{Dyson} of other 
conceivable life forms suggests that a more appropriate measure is that 
of the amount of information, $S_e$, effectively processed during the 
perception. 

The relevant information value, $S_e$ say, may be roughly estimated as 
the information capacity of the relevant number ${\cal N}_e$ of Everett 
type branch-channels (meaning the maximum of the Shannon entropy obtained 
by summing $-P\,{\rm log}\{P\}$ for a probability distribution $P$ over
the branch-channels) which will be given simply by
\be S_e={\rm log}\{ {\cal N}_e \} \, ,\label{3} \fe
(using a logarithm with base 2 for ordinary bit units). If it is supposed 
that the relevant branch-channels all have the same dimension, which will be 
given by the rank of the projection operator ${\bf e}$ associated with the 
perception, namely its trace ${\rm tr}\{{\bf e}\}$, then in a Hilbert space 
with dimension ${\rm tr}\{{\bf I}\}$ (where ${\bf I}$ is the identity 
operator) the required branch-channel number will be given by 
\be {\cal N}_e ={\rm tr}\{{\bf I}\}/{\rm tr}\{{\bf e}\}
 \, .\label{4} \fe
(It is to be noted that this ratio ${\cal N}_e$ is unaffected by the 
scaling up process that will occur if the Hilbert space for a local system 
is extended to the higher dimensional Hilbert space needed for describing 
a larger chunk of the universe. It is also to be noted that it is
multiplicative, so that the logarithm (\ref{3}) will be additive, when
perceptions by distinct individuals are merged to form the perception 
of a team or of an entire civilisation. ) 

Dyson estimated that the total entropy $Q_e$ generated during a human 
perception would be of the order of Avogadro's number, but that includes 
metabolic processes throughout the body, and so exceeds the Landauer 
entropy resulting~\cite{Vedral} from the relevant mental information 
$S_e$ by an enormous thermodynamic waste factor, $W_e=Q_e/S_e\gg 1$. 

The ansatz proposed here, namely
\be P\{{\cal O}_e\}\propto S_e\, ,\label{5}\fe
with
\be  S_e={\rm log}\big\{{\rm tr}\{{\bf I}\}\big\} -
{\rm log}\big\{{\rm tr}\{{\bf e}\}\big\}\, ,\label{6}\fe
may appropriately be designated by the term {\it entropic principle}
(paraphrasing the cruder anthropic principle that it supersedes). It is 
to be hoped that future investigations (of a neurological nature) may 
provide estimates of the absolute value of the relevant mental information 
$S_e$ that are more precise than the vague indication provided here (namely 
that in the adult human case it is very large compared with unity and very 
small compared with Avogadro's number) and that may be able to evaluate
the decrease that presumably occurs during a transition from a waking to
a dreaming state.  However the importance of such uncertainties (and of the 
even greater uncertainty concerning the proportionality factor in (\ref{5}), 
which is affected by the weighting that may be attributed to conceivable 
extraterrestrials) is diminished by the consideration that many 
applications require only relative probabilities, not absolute probability 
values.

Substitution of  (\ref{1}) and (\ref{5}) in (\ref{2}) leads to a 
prescription giving the probability of a perceptible eventuality
$e$ in the final form
\be P\{e\}\propto S_e\, {\rm tr}\{ {\bf P}\,{\bf e}\} \, ,\label{7}\fe
with $S_e$ given by (\ref{6}). The kind of approach encapsulated in this 
formula allows considerable latitude for adaptation  to satisfy idealogical 
desiderata of various alternative kinds, such as belief that the von-Neumann 
operator ${\bf P}$ should be obtained from a pure state function 
$\vert \Psi\rangle$ of the universe~\cite{HH, Vilenkin}, and prejudice for 
or against~\cite{Leslie} ontological prolificity. For example as formulated 
above, in terms of a single (though far from simple~\cite{Page}) act of 
perception, the interpretation presented here may not entirely satisfy 
those who demand that probabilities be defined in terms of frequencies. 
However that can easily be remedied by supposing that a very large number 
of perceptions is made (by one or many perceptors). So long as it does not 
affect the values of the distributions given by  $P\{{\cal O}_e\}$ and 
${\bf P}$, such a multi-perceptual reinterpretation makes no difference 
for practical (purely scientific) predictive purposes, though it might be 
slightly less satisfactory with respect to Ockham's ``razor'' criterion and 
much less satisfactory from the point of view of ontological economy.

The author wishes to thank Martin Rees for hosting the meeting at which 
this proposal was first informally mooted.

\end{document}